\documentclass[reprint]{revtex4-1}

\usepackage{amsmath}    
\usepackage{graphicx}   
\usepackage{verbatim}   
\usepackage{color}      
\usepackage{hyperref}   

\begin{document}

\title{Spontaneous valley polarization of interacting carriers in a monolayer semiconductor}

\author{J. Li$^{1}$, M. Goryca$^{1}$, N.~P. Wilson$^{2}$, A.~V. Stier$^1$, X. Xu$^{2}$, S.~A. Crooker$^{1}$}
\affiliation{$^1$National High Magnetic Field Laboratory, Los Alamos National Laboratory, Los Alamos, NM 87545}
\affiliation{$^2$Department of Physics, University of Washington, Seattle, WA 98195}

\date{\today}

\begin{abstract}
We report magneto-absorption spectroscopy of gated WSe$_2$ monolayers in high magnetic fields up to 60~T. When doped with a 2D Fermi sea of mobile holes, well-resolved sequences of optical transitions are observed in both $\sigma^\pm$ circular polarizations, which unambiguously and separately indicate the number of filled Landau levels (LLs) in both $K$ and $K'$ valleys. This reveals the interaction-enhanced valley Zeeman energy, which is found to be highly tunable with hole density $p$. We exploit this tunability to align the LLs in $K$ and $K'$, and find that the 2D hole gas becomes unstable against small changes in LL filling and can spontaneously valley-polarize.  These results cannot be understood within a single-particle picture, highlighting the importance of exchange interactions in determining the ground state of 2D carriers in monolayer semiconductors.
\end{abstract}

\maketitle

Electron-electron (\textit{e-e}) interactions underpin many interesting phenomena in 2D layers of mobile charges, including the fractional quantum Hall effect \cite{Stormer1999}, spin textures (skyrmions) \cite{Sondhi1993}, and quantum Hall ferromagnetism \cite{Girvin2000}. These phenomena arise from the Coulomb repulsion between charges, which in turn typically enhances the susceptibility of spin or related pseudospin (\textit{e.g.}, valley, layer, subband) degrees of freedom \cite{Girvin2000, Jungwirth2000, Ando1974, DasSarma2009}, and can even cause instabilities and spontaneous transitions to broken-symmetry phases \cite{Giuliani1985, Yarlagadda1991, Attaccalite2002, Wojs2002}. Such interactions have been studied in 2D electron and hole gases (2DEGs, 2DHGs) in conventional Si, GaAs, and AlAs semiconductors \cite{Koch1993, Daneshvar1997, Piazza1999, DePoortere2000, Brosig2000, Zhu2003, Gunawan2006} (and also in graphene \cite{Nomura2006, Young2012}), usually deep in the quantum regime at high magnetic fields $B$ where only a few Landau levels (LLs) are occupied.  Studies in tilted $B$ have proven indispensable in these materials \cite{Koch1993, Daneshvar1997, DePoortere2000, Brosig2000, Zhu2003}, because they provide a means to tune orbital (cyclotron) and spin (Zeeman) energies independently, thereby allowing to align LLs with different quantum numbers, so that \textit{e-e} interactions can manifest most clearly.

In the newer family of monolayer transition-metal dichalcogenide (TMD) semiconductors such as MoS$_2$ and WSe$_2$ \cite{Xiao2012, Xu2014, MakShan2016, Schaibley2016, Urbaszek2018}, advances in material quality have enabled high-mobility 2DEGs and 2DHGs \cite{Rhodes2019}. Owing to large carrier masses and reduced dielectric screening, \textit{e-e} interactions are anticipated to be strong, even at high carrier densities. Of particular interest, band extrema lie at the inequivalent $K$ and $K'$ points (valleys) of the Brillouin zone, providing exciting opportunities to study both spin \textit{and} valley degrees of freedom in doped monolayer systems. Indeed, recent transport \cite{Movva2017, Larentis2018, Pisoni2018, Lin2019}, optical \cite{Wang2017, Back2017, Wang2018, Smolenski2019, Liu2020, SuFeiShi2020}, and compressibility \cite{Gustafsson2018} studies revealed the expected \cite{Cai2013, Rose2013} spin- and valley-polarized LLs and related quantum oscillations in large $B$, and enhanced valley susceptibilities have been inferred. However, because spins in TMD monolayers are locked out-of-plane by strong spin-orbit coupling, tilted-$B$ methods cannot align LLs with different valley/spin index \cite{Movva2017, Larentis2018, Pisoni2018, Lin2019}.  To date this has limited studies of predicted \cite{Braz2018, Donk2018, Miserev2019, Roch2020} valley/spin instabilities and phase transitions arising from \textit{e-e} interactions.

Using a hole-doped WSe$_2$ monolayer, here we demonstrate and then exploit the density-tunable enhancement of the valley Zeeman energy to align LLs in the $K$ and $K'$ valleys.  Under these conditions, the 2DHG becomes unstable and exhibits spontaneous valley polarization. To observe this we measure absorption spectra in large magnetic fields to 60~T, and find well-resolved sequences of optical transitions in both $\sigma^+$ and $\sigma^-$ circular polarizations.  Due to the valley-specific optical selection rules, this allows to unambiguously and separately determine the number of filled LLs in \textit{each} valley for the first time. We identify up to eight valley-polarized LLs at low hole density $p$, which drops to five as $p$ increases, revealing a highly tunable valley susceptibility. By using this effect to tune LLs in the two valleys into near-alignment, we show that the disappearance and sudden reappearance of optical absorption in the $K'$ valley reveals the spontaneous transfer of holes to the $K$ valley, highlighting the key role of \textit{e-e} interactions in monolayer TMDs.

The experiment is depicted in Fig. 1(a). Using a dry stacking procedure, an exfoliated WSe$_2$ monolayer was sandwiched between hexagonal boron nitride slabs (hBN; $\sim$25~nm thick), and graphite flakes (2-4~nm thick) served as the contact and the top/bottom gates. The dual-gated device was placed over the 3.5~$\mu$m diameter core of a single-mode optical fiber having Ti/Au electrodes.  The device-on-fiber assembly ensured a rigid alignment of the optical path, mitigating  vibrations and drift. The assembly was mounted in the cold (4~K) bore of a 60~T pulsed magnet. Unpolarized white light from a xenon lamp was directed into the fiber and through the gated monolayer, and then through a thin circular polarizer before being redirected into a multi-mode collection fiber. The collected light was detected by a spectrometer and fast CCD detector.  Full spectra were continuously acquired every 0.5~ms throughout the magnetic field pulse.

Figure 1(b) shows a typical map of the absorption spectra versus gate voltage $V_g$ at $B$=0. Red features indicate absorption. Only the neutral (bare) exciton resonance $X^0$ is observed (at $\simeq$1.722~eV) when $-1.3$V$< V_g <$0V, where the monolayer contains no free carriers. However, when $V_g >0$, electrons populate the conduction bands, and lower-energy absorption from negatively-charged trions appears. Conversely, when $V_g < -1.3$~V, holes populate the valence bands and the positively-charged trion appears at lower energy ($\simeq$1.70~eV). Similar maps have been reported recently \cite{Wang2017, Liu2020, Courtade2017}; the narrow features observed here confirm the high quality of our sample-on-fiber assembly. These features can also be described as many-body exciton-polarons \cite{Back2017, Smolenski2019, Suris2001, Efimkin2017, Efimkin2018, Glazov2020}. That is, an electron-hole pair photoexcited into an existing Fermi sea will be dressed by interactions with the mobile carriers in the opposite valley, leading to distinct ``attractive'' and ``repulsive'' branches of the exciton-polaron quasiparticle. At low carrier densities, these branches equate with the lower-energy trion and higher-energy exciton, respectively \cite{Glazov2020}. As density increases, the repulsive branch blueshifts and fades as all the oscillator strength shifts to the lower-energy attractive branch.

We focus henceforth on $p$-type WSe$_2$ because it provides an especially simple realization of a two-valley system: The huge 450~meV spin-orbit splitting between the spin-up and -down valence bands in $K$ and $K'$ ensures that only the topmost band in each valley plays a role, even at the largest $p$ and $B$.

\begin{figure}[t]
\centering
\includegraphics[width=0.99\columnwidth]{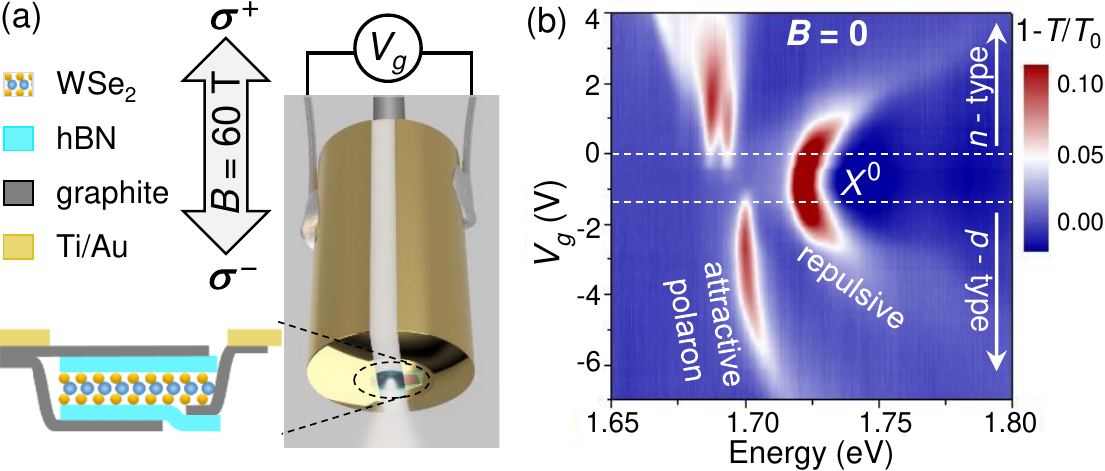}
\caption{\label{Fig1} (a) The dual-gated WSe$_2$ monolayer-on-fiber assembly used for absorption spectroscopy in pulsed magnetic fields $B$ to $\pm$60~T. (b) A $B$=0 map of the absorption spectra ($1- \frac{T}{T_0}$) vs. gate voltage $V_g$, at 4~K. $T$ and $T_0$ are transmission and reference spectra.}
\end{figure}

Essential to our main goal of controlling the LL alignment in the two valleys is the ability to measure --unambiguously and separately-- the number of filled LLs in \textit{each} valley. As direct reporters of these values we use the bare exciton transition (in $K$), and the attractive polaron transition (in $K'$), as described below. This approach exploits the optical selection rules in monolayer TMDs, whereby transitions in $K$ and $K'$ couple selectively to $\sigma^+$ and $\sigma^-$ circularly polarized light. We emphasize that optical studies are therefore distinct from electrical measurements of quantum oscillations \cite{Movva2017, Larentis2018, Pisoni2018, Lin2019, Gustafsson2018}, which reveal overall LL filling factors but do not \textit{a priori} indicate in which valley the various LLs reside. Importantly, we also rely on large $B$ to work in the well-resolved quantum regime, where only a few LLs are occupied (even at large $p$) and interactions are strongest \cite{Ando1974}, and where spectral signatures from different LLs are distinct.

\begin{figure*}[t]
\centering
\includegraphics[width=1.95\columnwidth]{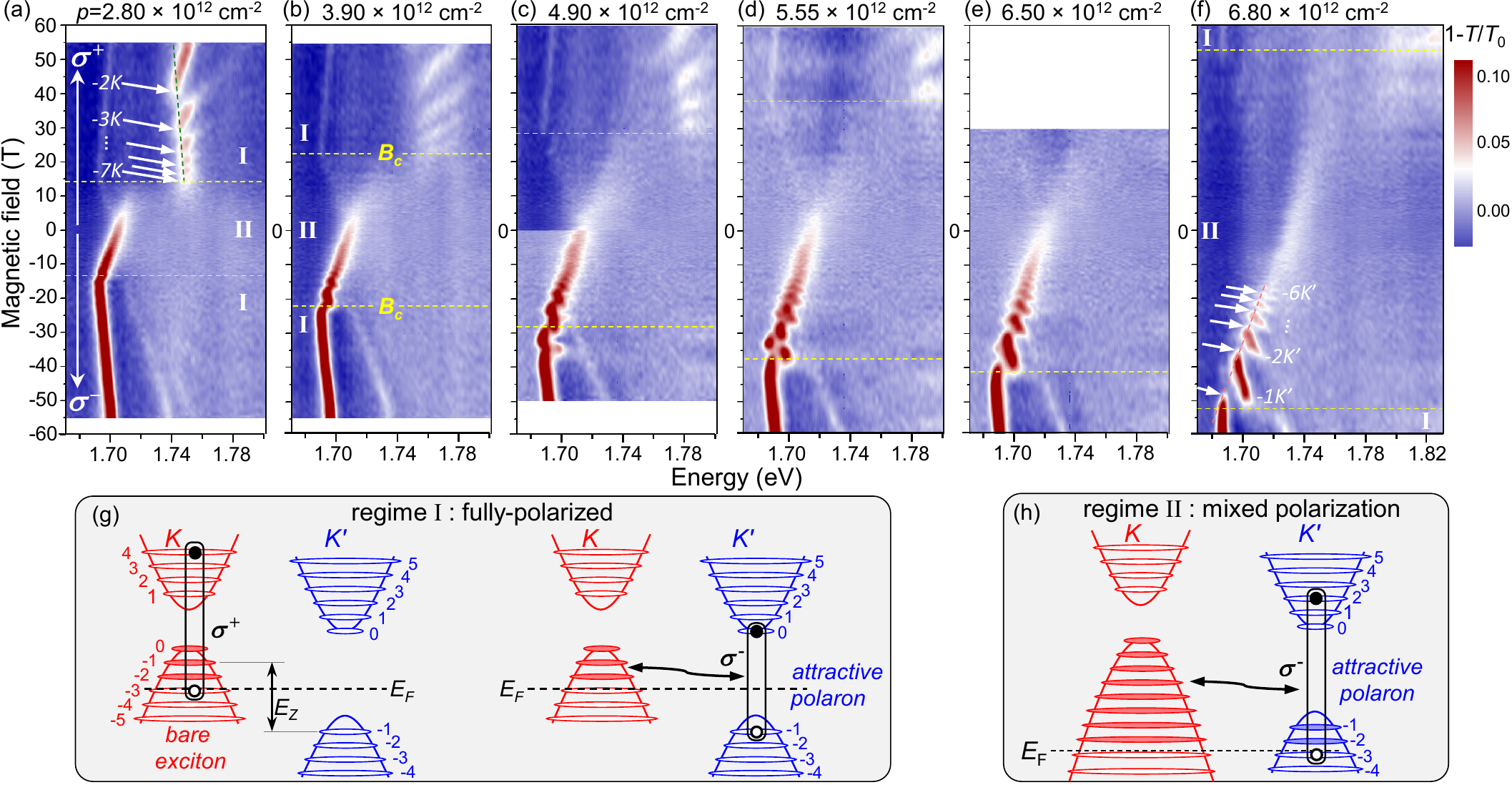}
\caption{\label{Fig2}(a-f) Maps of the $\sigma^\pm$ absorption spectra ($1-\frac{T}{T_0}$) vs. $B$ up to 60~T, at various hole densities $p$. $\sigma^+$ ($\sigma^-$) polarized spectra, \textit{i.e.}, from the $K(K')$ valley, are observed at $+B (-B)$. Horizontal lines separate the ``fully-polarized'' regime I ($|B| > B_c$) from the ``mixed'' regime II ($|B| < B_c$). Arrows show where the indicated LLs in $K$ and $K'$ become completely filled with holes. (g) Diagrams of the LLs \cite{Cai2013, Rose2013} and optical transitions in regime I (uninvolved bands are not drawn): The series of $X^0$ transitions (in $\sigma^+$) from the highest unfilled LL in $K$, and the attractive polaron transition (in $\sigma^-$) from the top LL in $K'$ (panels a and b show good examples).  (h) In regime II, the series of attractive polaron transitions from the highest unfilled LL in $K'$ (panels e and f show good examples). Spontaneous valley polarization occurs when $p$ is tuned to $5.5 \times 10^{12}$~cm$^{-2}$ (panel d).}
\end{figure*}

Figure 2 shows maps of the absorption spectra versus $B$, at selected $p$.  Optical transitions in the $K(K')$ valley are revealed in $\sigma^+ (\sigma^-)$ polarization, obtained in $+B (-B)$ \cite{Goryca2019}. Supplemental Fig. S1 shows maps at all measured $p$.  Each map is separated by dashed horizontal lines at a $p$-dependent crossover field $\pm B_c$, into two regimes: In the ``fully-polarized'' regime I ($|B|>B_c$), the 2DHG resides entirely in the $K$ valley; this occurs when the valley Zeeman splitting $E_Z$ exceeds the Fermi energy $E_F$. In $\sigma^+$, transitions are those of the bare exciton $X^0$ in $K$ (because there are no holes in $K'$ with which to form polarons). These $X^0$ transitions, depicted in Fig. 2(g), can be regarded as photoexcitation of an electron at $E_F$ (to the appropriate conduction band) and are clearly observed, \textit{e.g.}, in Figs. 2(a-c) as a series of discrete resonances at large $+B$. As $B$ decreases, a new resonance appears whenever the $-i^{th}$ LL in $K$ fills with holes and the Fermi level jumps to the next $-(i+1)^{th}$ LL. Note that this sequence ceases when $B$ falls below $B_c$, the field below which holes begin to also fill the $K'$ valley.  Counting the $X^0$ resonances at large $+B$ reveals how many LLs in $K$ are filled before any LLs in $K'$ are filled, from which we determine $E_Z$.  For example in Fig. 2(a) where $p$ is small, six resonances appear between +55~T and +10~T, and from maps at even lower $p$ (see Supplementary Fig. S1) we can track the filling down to the first ($0K$) level. This allows us to assign the resonances in Fig. 2(a) to the filling of the $-2K \rightarrow -7K$ levels, indicating that \textit{eight} LLs in $K$ are filled before the first LL in $K'$. Therefore $E_Z$ (=$2 g_h^* \mu_B B$) is much larger than the LL spacing (\textit{i.e.}, the cyclotron energy $\hbar eB/m_h$). Using a hole mass $m_h$=0.5~$m_0$ \cite{Gustafsson2018}, the effective hole g-factor $g_h^*$  must therefore lie between 12 and 14, which greatly exceeds its non-interacting value of $\approx 5.5$ \cite{Aivazian2015}.

Still in regime I, but now in $\sigma^-$ polarization (large negative $B$), the primary resonance is the attractive polaron.  As depicted in Fig. 2(g), it is formed by an exciton photoexcited from the uppermost $-1K'$ level, which couples to the 2DHG in $K$. It redshifts linearly with decreasing $|B|$ as the conduction and valence bands in $K'$ approach, until $|B|$ falls below $B_c$ and holes begin to also populate $K'$, whereupon a kink is observed as Pauli blocking effects commence, and the system enters regime II.

In the ``mixed'' regime II ($|B|<B_c$; see Fig. 2(h)), holes occupy LLs in \textit{both} valleys. Bare $X^0$ transitions therefore vanish. In $\sigma^-$, attractive polarons are excited from the Fermi level, which is now located within the ladder of LLs in $K'$. As $|B|$ decreases and these LLs fill with holes, a sequence of discrete absorption lines are observed [\textit{e.g.}, Figs. 2(e,f)], which can be regarded as attractive polarons originating from $-2K'$,..., $-iK'$. The arrows indicate where $-iK'$ fills completely, and absorption from the next (unfilled) LL commences.  \textit{Crucially, by tracking these resonances we determine which LLs in $K'$ are filled}.  Together with the filled LLs in $K$ obtained from the $X^0$ resonances in $\sigma^+$ (described above), we can reconstruct an accurate picture of the filled LLs in \textit{each} valley.  This is an essential step towards determining, and controlling, the alignment of LLs in $K$ and $K'$.

\begin{figure}[t]
\centering
\includegraphics[width=0.99\columnwidth]{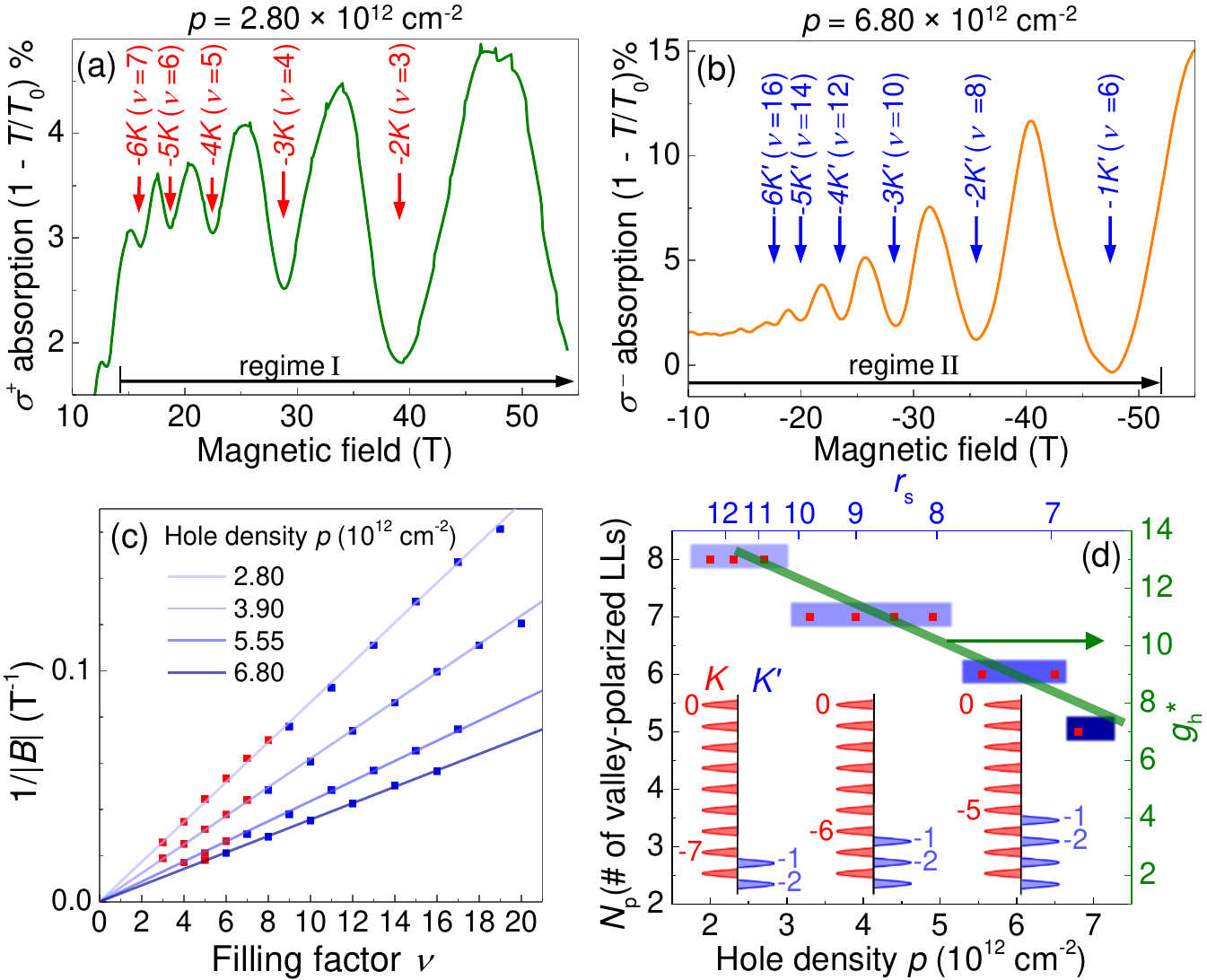}
\caption{\label{Fig3} (a,b) Absorption along the green and orange line-cuts drawn in Figs. 2a and 2f, respectively. SdH-like minima occur when the indicated LLs are completely filled.  Overall filling factors $\nu$ are also indicated. (c) $|B|^{-1}$ vs. $\nu$, at different $p$.  Red (blue) points are derived from $\sigma^+$ ($\sigma^-$) spectra; \textit{i.e.}, from LLs filled in $K$ ($K'$). (d) $N_p$, the number of valley-polarized LLs in $K$ (left axis) and corresponding $g_h^*$ (right) vs. $p$ and $r_s$. Diagrams show LL ordering at different $N_p$.}
\end{figure}

To quantify this picture, Figs. 3(a) and 3(b) show examples of the absorption along the indicated line-cuts through the $X^0$ resonances in Fig. 2(a) (regime I, $\sigma^+$), and through the attractive polaron resonances in Fig. 2(f) (regime II, $\sigma^-$), respectively.  Both show $1/B$-perodic Shubnikov-de Haas (SdH) type oscillations.  Minima occur when the indicated LL is filled, and correspond to the arrows shown on the absorption maps. In regime I, adjacent minima change the overall filling factor $\nu$ by one (because only $K$ is occupied), whereas adjacent minima in regime II change $\nu$ by two (because both valleys are occupied).

Importantly, Fig. 3(c) shows that for a given $p$, minima in \textit{both} $\sigma^\pm$ (red and blue points) can be combined and fit to the single line given by $\nu = hp/eB$, where $e/h=2.415 \times 10^{10}$~cm$^{-2}$T$^{-1}$ is the LL degeneracy. This confirms the enumeration of filled LLs and determines $p$. Moreover, these data also reveal $N_p$, the number of LLs in $K$ that must be filled before any LL in $K'$ is filled.  As discussed above, we determine $N_p$=8 when $p$ is small ($< 3 \times 10^{12}$~cm$^{-2}$), indicating a large $E_Z$ and enhanced $g_h^*$. However, as shown in Fig. 3(d), $N_p$ falls from 8$\rightarrow$5 as $p$ increases to $\approx$7$\times$10$^{12}$~cm$^{-2}$, revealing that $E_Z$ and therefore $g_h^*$ are \textit{strongly dependent} on $p$. The density-dependent valley susceptibility and ordering of the LLs, depicted in Fig. 3(d), is unambiguously observed here via optical spectroscopy for the first time. These results are consistent with recent reports of odd-even orderings of SdH oscillations \cite{Movva2017, Larentis2018, Pisoni2018, Lin2019, Gustafsson2018}, and in line with the expectation that \textit{e-e} interactions increase at small $p$ \cite{DasSarma2009, Zhu2003}.  Figure 3(d) also shows the dimensionless interaction parameter $r_s= m_h e^2/(\hbar^2 \kappa \sqrt{\pi p})$, which characterizes the ratio of Coulomb to kinetic (Fermi) energy \cite{DasSarma2009}, where $\kappa$=3 is the hBN dielectric constant.  $r_s \gg 1$ even at large $p$, which anticipates the important role of \textit{e-e} interactions.

\begin{figure}[t]
\centering
\includegraphics[width=.99\columnwidth]{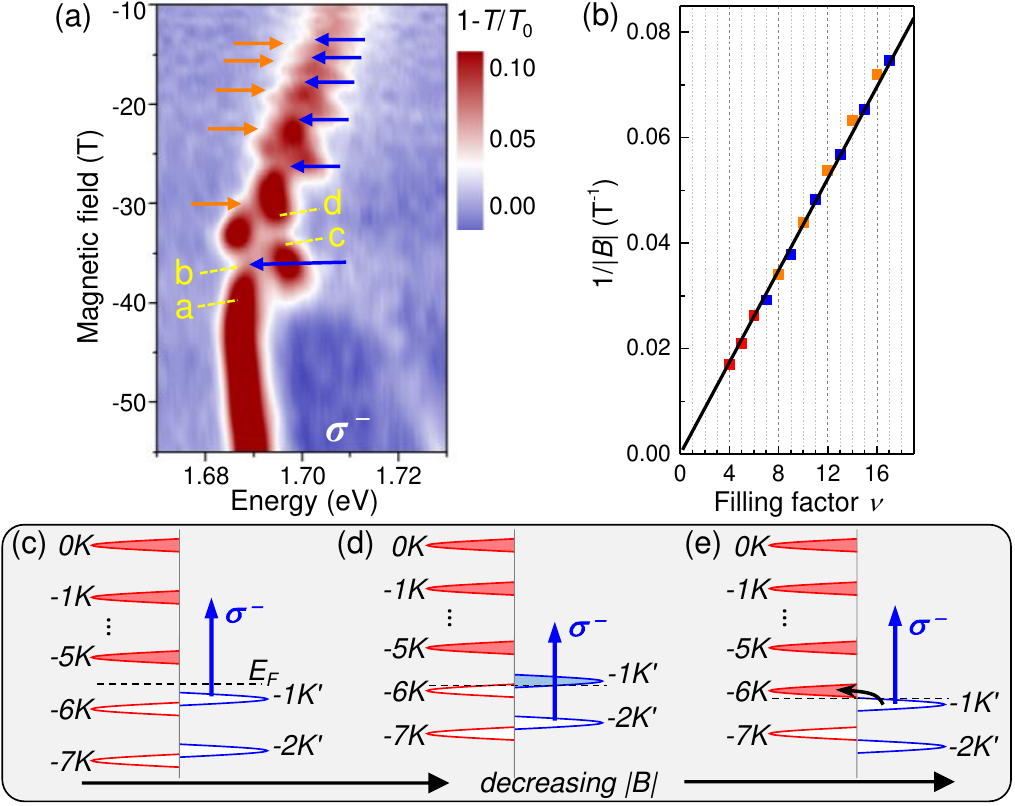}
\caption{\label{Fig4}(a) Expanded $\sigma^-$ absorption map when $p=5.55 \times 10^{12}$~cm$^{-2}$, for which $-6K$ and $-1K'$ LLs are nearly degenerate ($g_h^* \approx 10$). Attractive polaron absorption originates from the highest \textit{unfilled} LL in $K'$. The disappearance and reappearance of absorption as $|B|$ decreases from 40$\rightarrow$30~T reveals the filling and abrupt emptying of $-1K'$, due to spontaneous valley polarization.  Blue and orange arrows indicate the complete filling of LLs. (b) $|B|^{-1}$ vs. $\nu$. (c-e) Schematics of the $-1K'$ level filling and emptying as $|B|$ decreases.}
\end{figure}

Having established the interdependence between $g^*_h$, $\nu$, $B$, and $p$, we now focus on the central goal: aligning LLs in $K$ and $K'$ so that \textit{e-e} interactions can potentially drive instabilities in the valley pseudospin.  Historically, this was often achieved by tilting $B$ to tune LLs with different spin into alignment \cite{Girvin2000, Koch1993, Daneshvar1997, DePoortere2000, Brosig2000, Zhu2003}. In TMD monolayers where spins are locked out-of-plane, alternative approaches are required. We therefore exploit the dependence of $E_Z$ and $g_h^*$ on $p$. Figure 3(d) shows that LLs in $K$ and $K'$ will align only at certain $p$ where $E_Z$ is a multiple of the cyclotron energy (equivalently, when $N_p$ changes).  Using $m_h = 0.5 m_0$, this conveniently occurs when $g_h^*$ is an even integer. 

Instabilities will be most apparent at large $B$, where transitions from adjacent LLs are well-resolved spectrally. We therefore tune $p \approx 5 \times 10^{12}$~cm$^{-2}$, so that $g_h \approx 10$ and $B_c \approx 40$~T. And indeed, under these conditions, we find that signatures of instability and spontaneous valley polarization are observed in the expanded $\sigma^-$ map of Fig. 4(a). In marked contrast to maps at larger and smaller $p$ that show a smooth and systematic progression of resonances as the 2DHG enters the mixed regime [\textit{e.g.}, Figs. 2 (b,e)], Fig. 4(a) shows that as $|B|$ decreases from 40$\rightarrow$30~T, the $\sigma^-$ absorption jumps back and forth between discrete values, revealing the \textit{filling, sudden depopulation, and re-filling} of the uppermost $-1K'$ LL, indicating spontaneous valley polarization.

This unusual pattern can be understood as follows: Initially, LLs in $K$ and $K'$ are ordered as drawn in Fig. 4(c): $g_h^*$ is slightly less than 10, so that $-1K'$ resides slightly above $-6K$. At large $|B|$, the 2DHG is fully polarized and the only $\sigma^-$ absorption is the attractive polaron originating from (the unoccupied) $-1K'$.  As $|B|$ falls below $\sim$40~T, the 2DHG enters the mixed regime II as $-1K'$ begins to fill with holes (point ``a'' on the map). When it fills completely (point ``b''), absorption from $-1K'$ ceases and absorption from $-2K'$ commences, as described above and as drawn in Fig. 4(d). Crucially, however, at $\sim$34~T (point ``c''), absorption from $-2K'$ abruptly and unexpectedly ceases, and absorption from $-1K'$ \textit{reappears}.  This indicates that $-1K'$ has at least partially emptied, and that the holes have transferred to $-6K$, discontinuously lowering the net \textit{e-e} interaction energy by maximizing the total spin/valley polarization. This can be regarded \cite{Ando1974} as an abrupt jump in $g_h^*$ to a value slightly \textit{larger} than 10, which re-orders the LLs [see Fig. 4(e)]. As $|B|$ decreases further, $-1K'$ eventually  must fill again and its associated absorption ceases once again, while absorption from $-2K'$ re-commences (point ``d''). Thus, the two fields at which absorption from $-1K'$ cease ($\sim$35~T and $\sim$30~T) correspond to $\nu$=7 and $\nu$=8. Moreover, although not as clearly resolved, a similar sequence of disappearing and reappearing absorption transitions is observed at lower fields whenever the $-i^{th}$ LL in $K'$ (which is nearly degenerate with the $-(i + 5)^{th}$ LL in $K$) fills with holes. By counting all these transitions, all $\nu$ up to 17 can be fit as shown in Fig. 4(b). This instability, observed in $p$-type WSe$_2$ via its selective influence on the attractive exciton-polaron optical transitions, provides direct evidence for a spontaneous and discontinuous change in valley polarization in a monolayer TMD semiconductor, analogous to the transition to the quantum Hall ferromagnet state studied in conventional semiconductors \cite{Girvin2000, Giuliani1985, Yarlagadda1991, Attaccalite2002, Wojs2002, Koch1993, Piazza1999, DePoortere2000}. 

Based on Fig. 3(d), instabilities can also be expected when $N_p$ drops from $8 \rightarrow 7$ ($p \approx 3 \times 10^{12}$). However, at this point $B_c \simeq 18$~T, which is not sufficient to spectrally resolve adjacent LLs in this sample -- although the data do suggest an anomaly (Supplemental Fig. S2). Similarly, instabilities may occur at larger $p \approx 7 \times 10^{12}$ (when $N_p$ drops from $6 \rightarrow 5$), but we did not observe it, possibly because $r_s$ is too small at this larger $p$ for \textit{e-e} interactions to dominate. 

These results indicate that the valence bands of monolayer TMDs represent a rich platform from which to study emergent phases arising from \textit{e-e} interactions.  We emphasize that the ability to unambiguously detect abrupt changes in the 2DHG valley polarization derives from the utility of polarized optical methods, which complement electrical measurements through their ability to \textit{separately} distinguish carriers in each valley.  Recent electrical measurements of SdH oscillations in various doped TMD monolayers have shown non-vanishing SdH minima when LLs are believed to align \cite{Pisoni2018, Lin2019, Gustafsson2018}, which is (at least) consistent with hybridization and anticrossing of degenerate LLs in $K$ and $K'$, and may also be consistent with a discontinuous jump in the LL ordering \cite{Koch1993}. Our polarized optical studies support the latter scenario. However, whether the valley instability that we clearly observe at $\approx$35~T represents an true first-order phase transition remains an open question; careful transport studies in this field and doping regime --and in particular the presence or absence of hysteretic resistance \cite{Piazza1999, DePoortere2000}-- could answer this question.

We thank Wang Yao, Jun Zhu, Bernhard Urbaszek, and Xavier Marie for helpful discussions. Work at the NHMFL was supported by the Los Alamos LDRD program and the DOE BES ‘Science of 100 T’ program. The NHMFL is supported by National Science Foundation (NSF) DMR-1644779, the State of Florida, and the U.S. Department of Energy (DOE). Work at the University of Washington was supported by the DOE Basic Energy Sciences, Materials Sciences and Engineering Division (Grant No. DE-SC0018171).

\newpage

\begin{figure*}[!htbp]
\centering
\includegraphics[width=2.0\columnwidth]{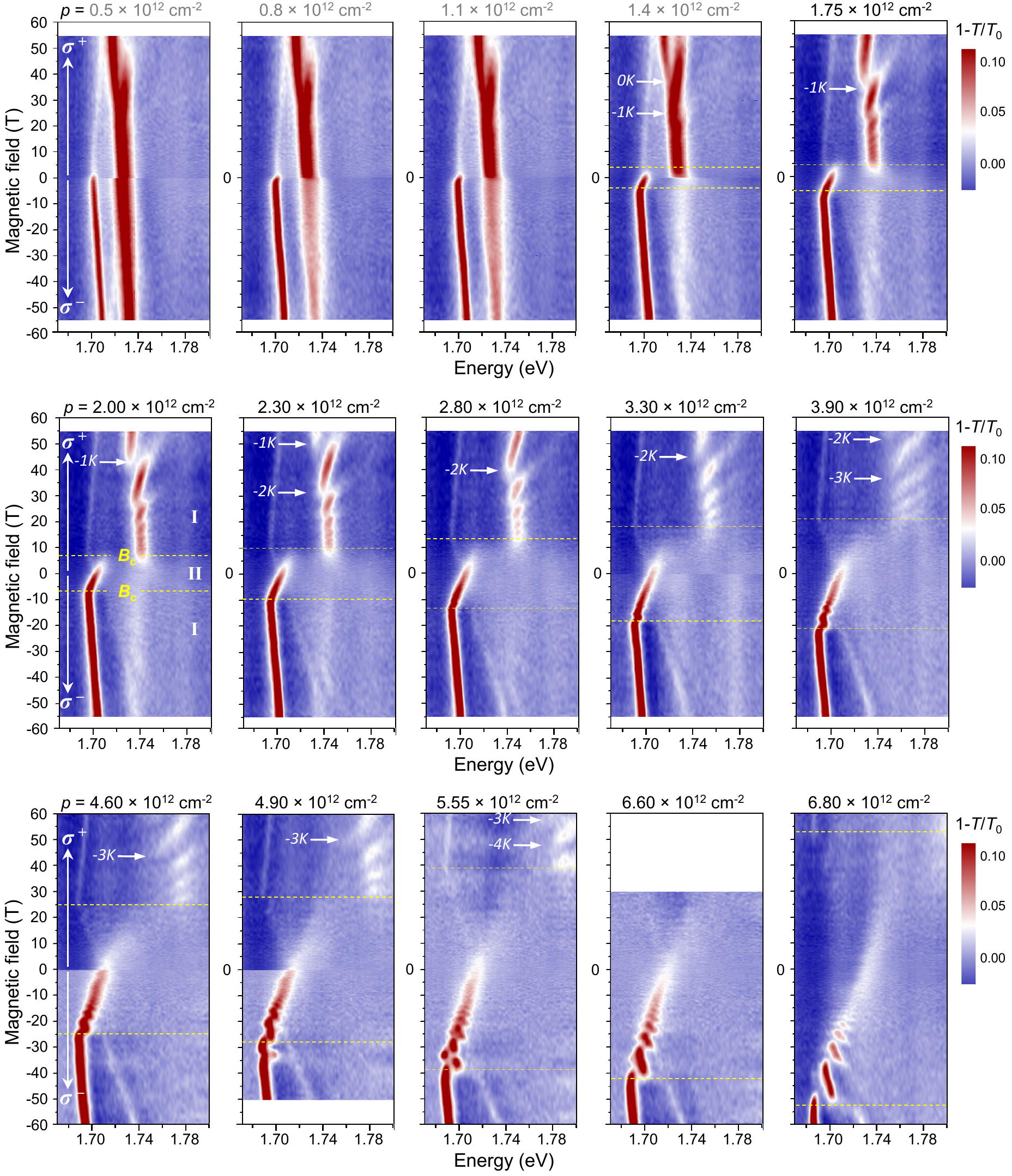}
\caption{\label{FigS1}\textbf{Supplemental Fig. S1}: Maps of the $\sigma^\pm$ absorption spectra ($1-T/T_0$) vs. $B$ up to $\pm 60$ T for all measured hole densities $p$. Densities $p\leq 1.4 \times 10^{12}$ cm$^{-2}$ are estimated by linear extrapolation of $p$ vs. $V_g$ relation. $\sigma^+$($\sigma^-$) polarized spectra, $i.e.$, from the $K$($K^\prime$) valley, are observed at $+B$($-B$). Horizontal lines separate the ``fully-polarized" regime I ($|B|>B_c$) from the ``mixed" regime II ($|B|<B_c$). As in Fig. 2 of the main text, arrows show where the indicated LL in the $K$ valley is completely filled with holes. At low $p$, we can track the filling down to the lowest $0K$ level.}
\end{figure*}

\begin{figure*}[!htbp]
\centering
\includegraphics[width=1.0\columnwidth]{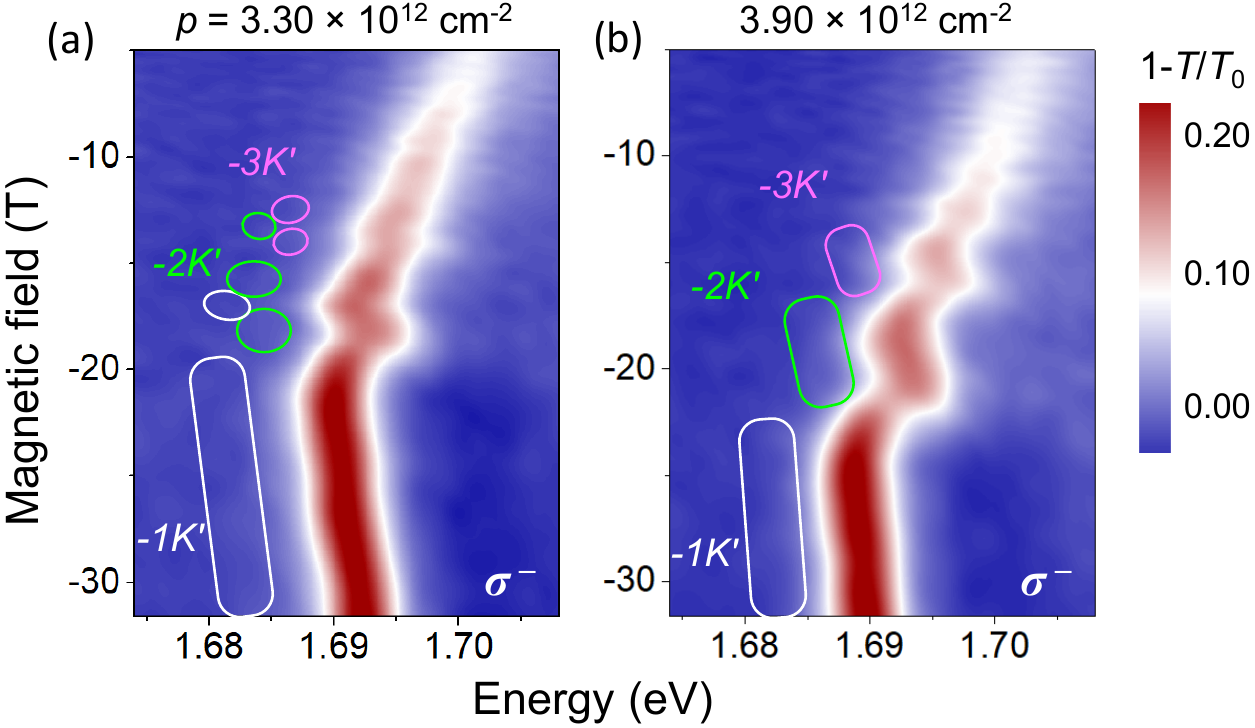}
\caption{\label{FigS2}\textbf{Supplemental Fig. S2}: Possible indications of the additional valley polarization instability that is expected at lower hole density ($p \sim 3 \times 10^{12}$ cm$^{-2}$), when $N_p$ drops from 8 to 7. Expanded $\sigma^-$ absorption maps at (a) $p = 3.30 \times 10^{12}$ cm$^{-2}$ and (b) $3.90 \times 10^{12}$ cm$^{-2}$. In marked contrast to the map on the right, which shows a smooth and systematic progression of the resonances as the 2DHG sequentially fills the LLs in the $K^\prime$ valley, the absorption line in the map on the left appears to more abruptly jump back and forth between discrete values (analogous to the instability shown in Fig. 4 of the main text). This instability is expected to occur when the $-1K^\prime$ level in the $K^\prime$ valley aligns with the $-7K$ level in the $K$ valley ($i.e.$, when $p \sim 3 \times 10^{12}$ cm$^{-2}$, and $N_p$ changes from 8 to 7). However, at this lower carrier density the $-1K^\prime$ level fills at around $B_c = 18$ T, and at this lower field the LLs are not sufficiently separated in energy to unambiguously resolve them in optical spectra. The oval shape drawings are guides to the eye of the $\sigma^-$ absorption features. White, green and magenta color represents attractive polaron absorption from the $-1K^\prime$, $-2K^\prime$, and $-3K^\prime$ LLs, respectively.}
\end{figure*}

\begin{figure*}[t]
\centering
\includegraphics[width=1.0\columnwidth]{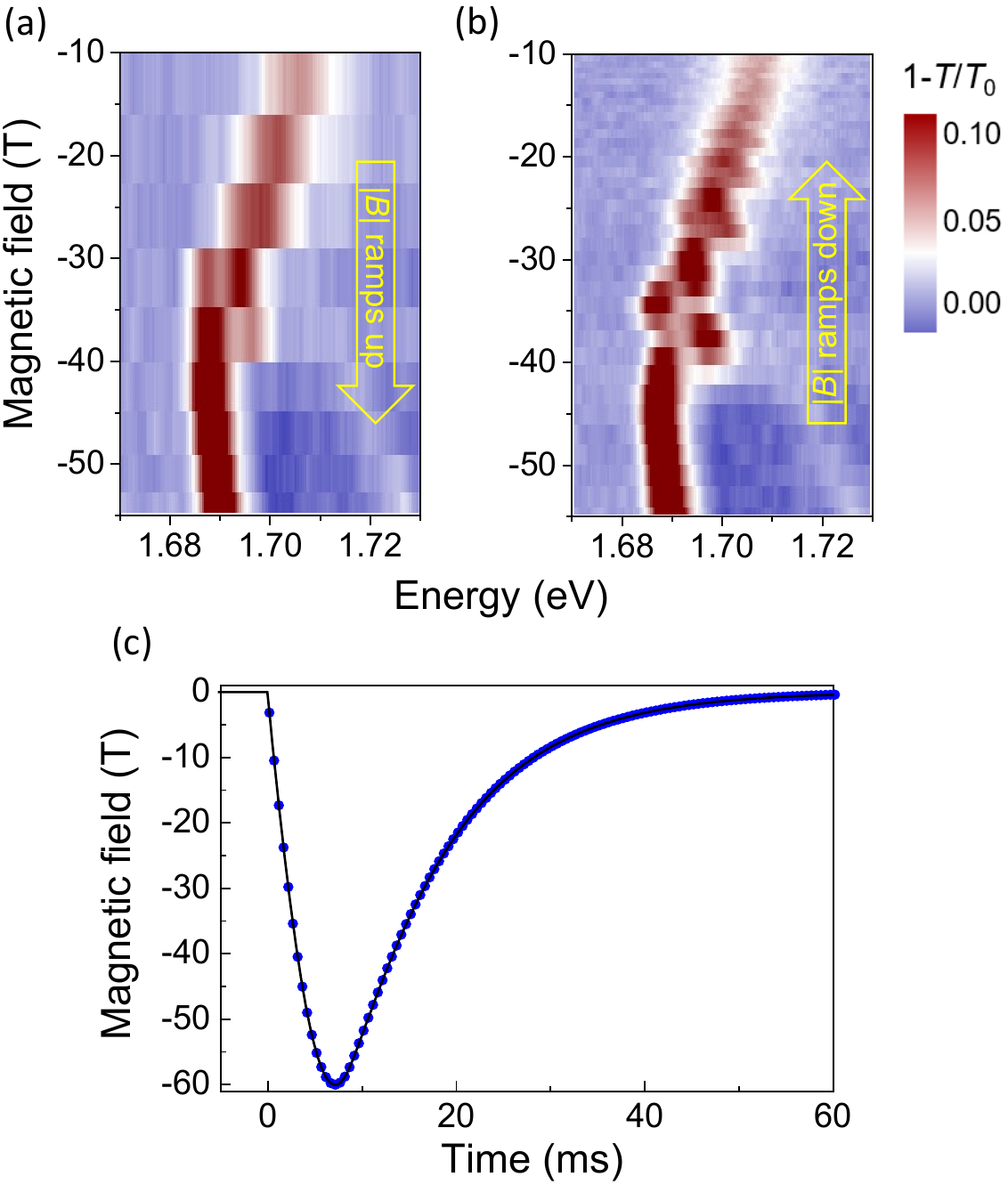}
\caption{\label{FigS3} \textbf{Supplemental Fig. S3}: Comparison of the raw (unsmoothed) $\sigma^-$ absorption maps at $p = 5.55 \times 10^{12}$ cm$^{-2}$, taken during a single magnetic field pulse when (a) $B$ sweeps quickly up to peak field (-60 T) and (b) when $B$ sweeps more slowly back down to zero field. Data shown in the main text is always acquired during the downsweep, where $B$ changes much less during the 0.5 ms spectral acquisition time (which is the fastest acquisition time of our CCD camera). (c) Field profile of the capacitor driven pulsed magnet, which has a fast upsweep and slow downsweep. Points show the average field at which each spectra was taken. The field is changing too rapidly during the upsweep to spectrally resolve the jumps in the absorption lines that are clearly resolved during the downsweep, precluding any quantitative analysis of possible hysteretic effects associated with spontaneous valley polarization.}
\end{figure*}

\end{document}